\RequirePackage{fix-cm}
\documentclass[twocolumn,epjc3]{svjour3}  
\smartqed  
\RequirePackage{graphicx}
\usepackage{amsmath}
\usepackage{amssymb}
\usepackage{hyperref}
\journalname{Eur. Phys. J. C}
\begin{document}

\title{Energy-Dependent Light Quenching in CaWO$_4$ Crystals at mK Temperatures
}


\author{R. Strauss \thanksref{1,2,e1} \and G. Angloher \thanksref{2} \and A. Bento \thanksref{3} \and C. Bucci \thanksref{4}\and L. Canonica \thanksref{4} \and W. Carli \thanksref{9} \and C. Ciemniak \thanksref{1} \and  A. Erb \thanksref{1,5} \and F.v.\,\,Feilitzsch \thanksref{1} \and P.\,\,Gorla \thanksref{4} \and A. G\"utlein \thanksref{1} \and D. Hauff \thanksref{2}  \and D. Hellgartner \thanksref{1}  \and J. Jochum \thanksref{6} \and H. Kraus \thanksref{7} \and J.-C. Lanfranchi \thanksref{1} \and J. Loebell \thanksref{6} \and A. M\"unster \thanksref{1} \and F.\,\,Petricca \thanksref{2} \and W. Potzel \thanksref{1} \and F. Pr\"obst \thanksref{2} \and F. Reindl \thanksref{2} \and S. Roth \thanksref{1} \and K. Rottler \thanksref{6} \and C. Sailer \thanksref{6} \and K. Sch\"affner \thanksref{4} \and J.\,\,Schieck \thanksref{8} \and S. Scholl \thanksref{6} \and S. Sch\"onert \thanksref{1} \and W. Seidel \thanksref{2} \and M.v.\,\,Sivers \thanksref{1} \and L. Stodolsky \thanksref{2} \and C. Strandhagen \thanksref{6} \and A.\,\,Tanzke \thanksref{2} \and M. Uffinger \thanksref{6} \and A. Ulrich \thanksref{1} \and I. Usherov \thanksref{6} \and S. Wawoczny \thanksref{1} \and M. Willers \thanksref{1} \and M. W\"ustrich \thanksref{2} \and A. Z\"oller \thanksref{1} 
}

\thankstext{e1}{e-mail: strauss@mpp.mpg.de}

\institute{Physik-Department, Technische Universit\"at M\"unchen \label{1}, D-85748 Garching, Germany \and Max-Planck-Institut f\"ur Physik,  D-80805 M\"unchen, Germany \label{2} \and CIUC, Departamento de Fisica, Universidade de Coimbra, P3004 516 Coimbra, Portugal \label{3}\and INFN, Laboratori Nazionali del Gran Sasso, I-67010 Assergi, Italy \label{4} \and Walther-Mei\ss ner-Institut f\"ur Tieftemperaturforschung,  D-85748 Garching, Germany \label{5} \and Physikalisches Institut, Eberhard-Karls-Universit\"at T\"ubingen,   D-72076 T\"ubingen, Germany \label{6} \and Department of Physics, University of Oxford, Oxford OX1 3RH, United Kingdom \label{7} \and Institut f\"ur Hochenergiephysik der \"Osterreichischen Akademie der Wissenschaften, A-1050 Wien, Austria \label{8} \and Maier-Leibnitz-Laboratorium, Ludwig-Maximilians-Universit\"at M\"unchen,   D-85748 Garching, Germany \label{9}
}

\date{Received: date / Accepted: date}

\maketitle

\begin{abstract}
Scintillating CaWO$_4$ single crystals are a promising multi-element target for rare-event searches and are currently used in the direct Dark Matter experiment CRESST (Cryogenic Rare Event Search with Superconducting Thermometers). The relative light output of different particle interactions in CaWO$_4$ is quantified by Quenching Factors (QFs). These are essential for an active background discrimination and the identification of a possible signal induced by weakly interacting massive particles (WIMPs). We present the first precise measurements of the QFs of O, Ca and W at mK temperatures by irradiating a cryogenic detector with a fast neutron beam. A clear energy dependence of the QF of O and, less pronounced, of Ca was observed for the first time. Furthermore, in CRESST neutron-calibration data a variation of the QFs  among different CaWO$_4$ single crystals was found. For typical  CRESST detectors the QFs in the region-of-interest (10-40\,keV) are $QF_O^{ROI}=(11.2{\pm}0.5)$\%, $QF_{Ca}^{ROI}=(5.94{\pm}0.49)$\% and $QF_W^{ROI}=(1.72{\pm}0.21)$\%. The latest CRESST data (run32) is reanalyzed using these fundamentally new results on light quenching in CaWO$_4$ having  moderate influence on the WIMP analysis. Their relevance for future CRESST runs and  for the clarification of previously published results of direct Dark Matter experiments is emphasized.
\end{abstract}
\section{Introduction}
Rare-event searches for Dark Matter (DM) in the form of weakly interacting massive particles (WIMPs) \cite{Bertone:2004pz,Jungman:1995df}  have reached impressive sensitivities during the last decade \cite{Cushman:2013zza}. Well motivated WIMP candidates with \linebreak masses $m_\chi$ between a few GeV/$c^2$ and a few TeV/$c^2$ might be detectable via nuclear recoils of  few keV in terrestrial experiments \cite{Lewin:1995rx}. While the DAMA/LIBRA \cite{Bernabei:2010mq}, and recently the CoGeNT \cite{Aalseth:2010vx}, CRESST \cite{Angloher:2012vn}, and the CDMS(Si) \cite{PhysRevLett.111.251301} experiments observed excess signals that might be interpreted as induced by DM particles with  $m_\chi{\sim10}$\,GeV/$c^2$ at WIMP-nucleon cross-sections \linebreak of  ${\sim}10^{-4}$\,pb, this scenario is  ruled out by the LUX \cite{Akerib:2013tjd} and XENON100 \cite{Aprile:2012nq} experiments, and  almost excluded  by the CDMS(Ge) \cite{Ahmed:2009zw,Ahmed:2010wy}, the EDELWEISS  \cite{Armengaud:2011cy,Armengaud:2012pfa} and the SuperCDMS \cite{Agnese:2014aze} experiments. It is strongly disfavoured by accelerator constraints \cite{ATLAS:2012ky,Chatrchyan:2012me} and in mild tension with an extended analysis \cite{PhysRevD.85.021301} of published CRESST data \cite{Angloher2009270}.
\section{Dark Matter search with CRESST}
The CRESST  experiment \cite{Angloher:2012vn} employs scintillating \linebreak CaWO$_4$ crystals  \cite{edison,PhysRevB.75.184308} as a multi-element target material. The key feature of a CRESST detector module is the simultaneous measurement of  the recoil energy $E_r$ by a particle interaction in the crystal (operated as cryogenic calorimeter at mK temperatures \cite{Probst:1995fk}) and the corresponding scintillation-light energy $E_l$ by a separate cryogenic light absorber. Since the relative light yield $LY{=}E_l/E_r$ is reduced for highly ionizing particles compared to electron recoils (commonly referred to as quenching) nuclear-recoil events can be discriminated from e$^-$/$\gamma$ and $\alpha$ backgrounds. The phenomenological Birks model \cite{birks1964theory} predicts this quenching effect to be stronger the higher the mass number $A$ of the recoiling ion, which allows to distinguish, in general, between  O ($A{\approx}16$), Ca ($A{\approx}40$) and W ($A{\approx}184$) recoils.   The expected WIMP-recoil spectrum - assuming coherent scattering - is completely dominated by W-scatters for $m_\chi\,\gtrsim 20$\,GeV/c$^2$. However, the light targets O and Ca make CRESST detectors particularly sensitive to low-mass WIMPs of $1$\,GeV$\,\lesssim\, m_\chi\,\lesssim$\,20\,GeV. Furthermore, the knowledge of the recoil composition of O, Ca and W allows a test of the assumed $A^2$-dependence of the  spin-independent  WIMP-nucleon cross-section \cite{Jungman:1995df}. In addition, background neutrons, which are mainly visible as O-scatters (from kinematics \cite{scholl_paper}), can be  discriminated statistically.
\section{Quenching Factors (QFs)}
The mean LY of e$^-$/$\gamma$ events ($LY_{\gamma}$) is energy \linebreak dependent and phenomenologically parametrized as\linebreak  ${LY_{\gamma}(E_r) = (p_0+p_1E_r) (1 - p_2\exp(-E_r/p_3))}$ \cite{strauss_PhD}. By \linebreak convention, $LY_{\gamma}$(122\,keV) is normalized to unity. The parameters $p_{0}$, $p_1$, $p_2$ and $p_3$ are derived from a maximum likelihood (ML) fit  for every detector module individually. For the module used in this work the fit yields:  $p_0\,{=}\,1.07$, $p_1\,{=}\,{-}1.40\cdot 10^{-5}$\,keV$^{-1}$, $p_2\,{=}\,6.94\cdot 10^{-2}$ and $p_3\,{=}\,147$\,keV (errors are negligible for the following analysis). The exponential decrease towards lower recoil energies (quantified by $p_2$ and $p_3$) accounts for the scintillator non-proportionality \cite{Lang:2009uh}. The Quenching Factor (QF) of a nucleus $x$ - in general energy dependent - is defined as  $QF_x(E_r)=LY_x(E_r)/LY_{\gamma,norm}$ where $LY_x$ is the mean LY of a nuclear recoil x and $LY_{\gamma,norm}$ is a detector-specific normalization factor which corresponds to the LY of e$^{-}$/$\gamma$ events. By convention,\linebreak $LY_{\gamma,norm}\,{=}\,LY_{\gamma}(E_r)/(1 - p_2\exp(-E_r/p_3))\,{\approx}\,p_0$ is used for the analysis since the scintillator non-proportionality is not observed for nuclear recoils and $p1\,\ll\,1$. For typical CRESST detector modules, the uncertainties in energy and LY are well described by gaussians \cite{Angloher:2012vn} consistent with photon-counting statistics in the energy range considered in this work.\\
Since the  resolution of light-detectors operated in the\linebreak CRESST setup at present is not sufficient to disentangle O, Ca and W recoils unambiguously, dedicated experiments to measure the QFs of CaWO$_4$ are necessary. Earlier attempts yield inconclusive results, in particular for the value of $QF_W$ \cite{Jagemann:2006sx,Ninkovic:2006xy,Bavykina:2007ze}. \\
\section{The neutron-scattering facility}
\subsection{Experimental setup}
\begin{figure}
\centering
\includegraphics[width=0.4\textwidth]{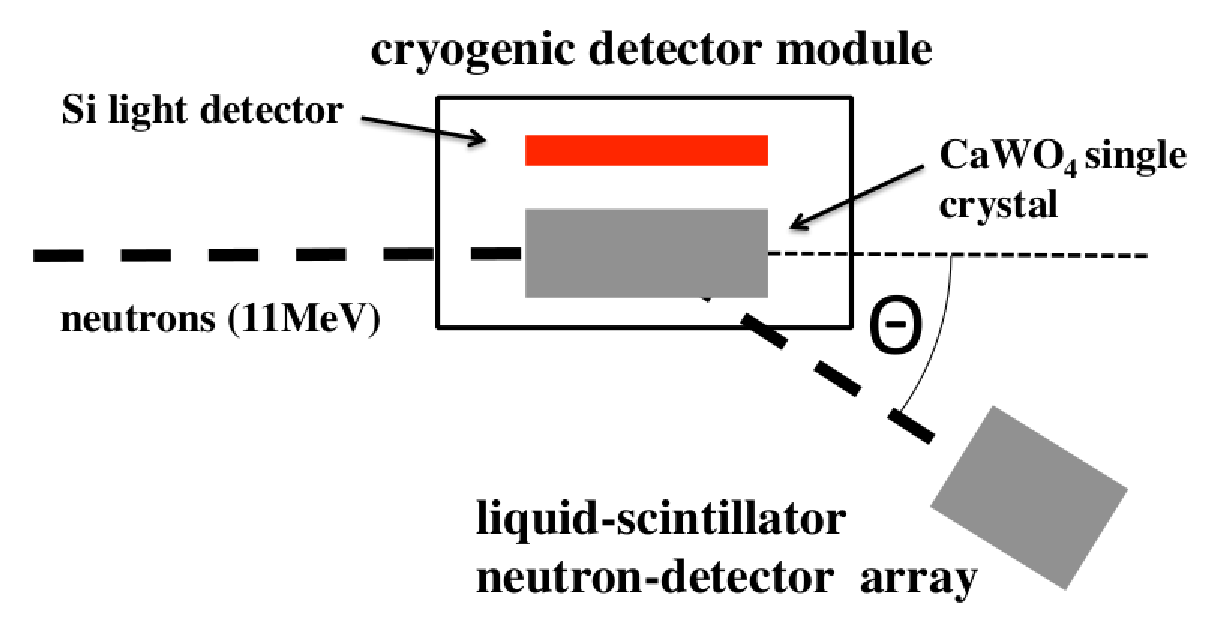}
\caption{Schematic experimental setup of the neutron-scattering facility. Neutrons produced by the accelerator are scattered off a CRESST-like detector module (operated at 20\,mK) and tagged in liquid-scintillator neutron detectors at a fixed scattering angle $\Theta$.} 
\label{fig:setupMLL}
\end{figure}

At the accelerator of the Maier-Leibnitz-Laboratorium \linebreak (MLL) in Garching a dedicated neutron-scattering facility for precision measurements of QFs at mK temperatures was set up (see Fig.\,\ref{fig:setupMLL}). A pulsed $^{11}$B beam  of ${\sim}65$\,MeV in bunches of 2-3\,ns (FWHM) produces monoenergetic neutrons of ${\sim}11$\,MeV  via the nuclear reaction p($^{11}$B,n)$^{11}$C in a pressurized H$_2$ target \cite{Jagemann2005245}. These neutrons are irradiated onto a CRESST-like detector module consisting of a ${\sim}$10\,g cylindrical CaWO$_4$ single crystal (20\,mm in diameter, 5\,mm in height) and a separated Si light absorber (20\,mm in diameter, 500\,$\mu$m thick) \cite{Strauss:2012fk}. Both are operated as cryogenic detectors in a dilution refrigerator at ${\sim}20\,$mK \cite{Lanfranchi20091405}. Undergoing elastic (single) nuclear scattering in CaWO$_4$ the neutrons are tagged at a fixed scattering angle $\Theta$ in an array of 40 liquid-scintillator (EJ301) detectors which allow fast timing (${\sim}2$\,ns) and n/$\gamma$ discrimination.
\subsection{Working principle}
Depending upon which of the three nuclei is hit a distinct amount of energy is deposited by the neutron in the crystal. Triple-coincidences between (1) a $^{11}$B pulse on the H$_2$ target, (2) a neutron pulse in a liquid-scintillator detector and (3) a nuclear-recoil event in the CaWO$_4$ crystal can be extracted from the data set. A neutron time-of-flight (TOF) measurement between neutron production and detection combined with a precise phononic measurement of the energy deposition in the crystal (resolution ${\sim}1$\,keV (FWHM)) allows an identification of the recoiling nucleus.  To derive the individual QF the corresponding scintillation-light output is  measured simultaneously by the light detector. Since the onset uncertainty of cryodetector pulses  is large (${\sim}5\,\mu$s) compared to typical neutron TOFs (${\sim}$50\,ns) an offline coincidence analysis has to be performed \cite{strauss_PhD}.
\subsection{Measurements and results for $QF_W$}
\begin{figure}
\includegraphics[width=0.48\textwidth]{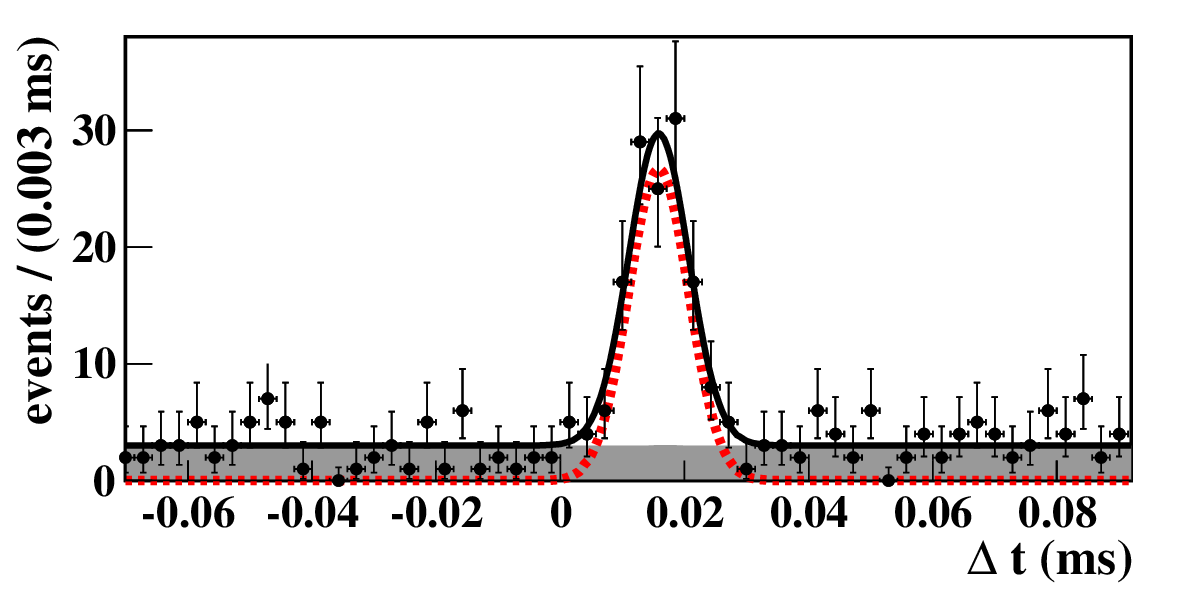} 
\caption{Histogram of the time difference $\Delta t$ between neutron events with the correct TOF and the closest W-recoil in the CaWO$_4$ crystal ($E_r\,{=}\,100\,{\pm}\,20$\,keV). A fit to the distribution (solid black line) including a constant for the accidental background (shaded area) and a gaussian for the triple-coincidences on W (dashed red line) is shown. 158 W-scatters are identified with a signal-to-background ratio of ${\sim}\,7:1$. }
\label{fig:timing} 
\end{figure}

\begin{figure}
\includegraphics[width=0.48\textwidth]{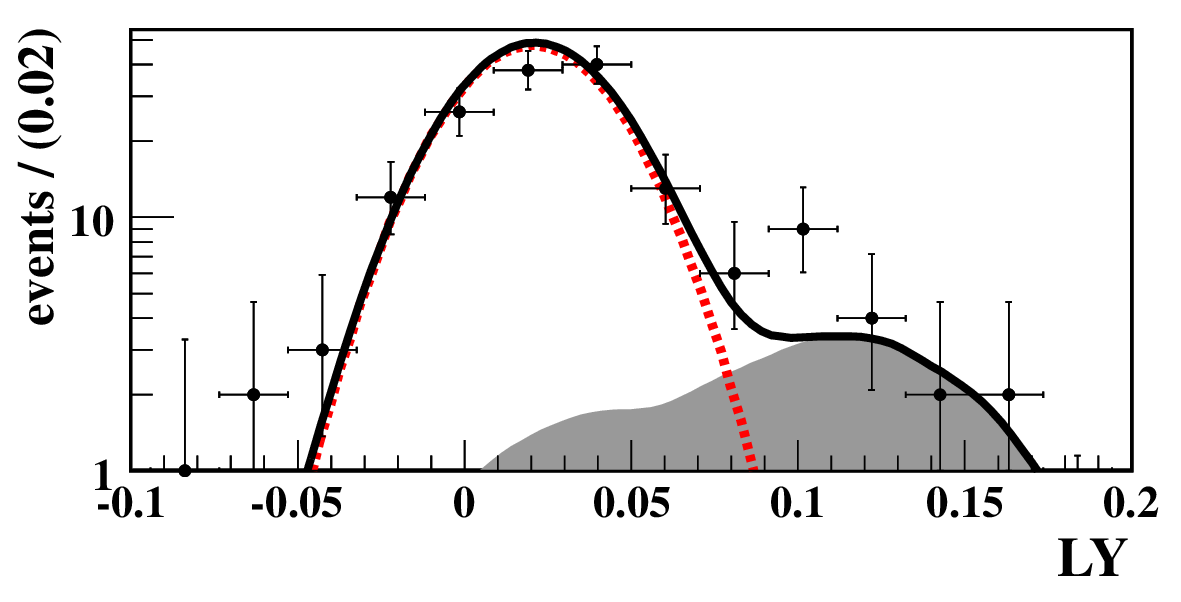} 
\caption{LY histogram of the 158 events identified as triple-coincidences on W. A fit to the distribution (solid black line) is shown which includes a gaussian (dashed red line) accounting for W-scatters and the background-pdf (shaded area)  describing accidental coincidences.  The simultaneous ML fit including the timing distribution  yields $QF_W\,{=}\,(1.96\,{\pm}\,0.22)$\%. }
\label{fig:LY}
\end{figure}

The experiment was optimized for the measurement of $QF_W$ \cite{strauss_PhD,Strauss:2014aa}. To enhance the number of W-scatters a scattering angle of ${\Theta\,{=}\,80^\circ}$ was chosen due to scattering kinematics \cite{Jagemann:2006sx}. For this specific angle, the expected  recoil energy of triple-coincident events is  ${\sim}\,100$\,keV for W, ${\sim}\,450$\,keV for Ca, and ${\sim}\,1.1$\,MeV for O. In  ${\sim}\,3$ weeks of beam time a total of  ${\sim}\,10^8$  cryodetector pulses  were recorded.  Fig.\,\ref{fig:timing} shows the time difference $\Delta t$ between neutron events  with the correct TOF identified in one of the liquid-scintillator detectors and the closest W-recoil (in time) in the CaWO$_4$ crystal ($E_r=100{\pm}20$\,keV). A gaussian peak of  triple-coincidences on W (dashed red line) at $\Delta t\,{\approx}\,0.016$\,ms and a width of $\sigma_t\,{\approx}\,4.8\,\mu$s (onset resolution of the cryodetector) is observed above a background due to accidental coincidences uniformly distributed in time (shaded area). Within the $2\sigma$-bounds of the peak 158 W-scatters are identified with a signal-to-background (S/B) ratio of ${\sim}\,7:1$.\\
Fig.\,\ref{fig:LY} shows the LY distrubution of these events in a histogram (black dots). The mean LY of the extracted W-scatters is found at a lower value compared to the mean LY of all nuclear recoils, i.e., the (overlapping) contributions of O, Ca and W if no coincidence measurement is involved. The accidental coincidences have a LY-distribution equal to that which is modelled by a probability-density function (background-pdf) \cite{strauss_PhD}. A simultaneous maximum-likelihood (ML) fit is performed including (1) the timing distribution which fixes the S/B ratio and the number of identified W-events, and (2) the LY distribution described by a gaussian (W-events) and the background-pdf.  The final results are  $LY_W\,{=}\,0.0208\,{\pm}\,0.0024$ and  $QF_W\,{=}\,(1.96\,{\pm}\,0.22)$\%, correspondingly (errors are dominated by statistics). Fig.\,\ref{fig:LY} shows the fit to the LY distribution by the gaussian (dashed red line) and the background-pdf (shaded area).\\ 
\section{Energy-dependent QF analysis}
\subsection{Principle of analysis}
For the measurement of $QF_{Ca}$ and $QF_O$ no coincidence signals are necessary, instead, an analysis of  the nuclear-recoil data alone is sufficient. Also for this analysis the neutron data obtained at the scattering facility were used. 
Commonly CRESST data is displayed in the energy-LY plane \cite{Angloher:2012vn} giving rise to nearly horizontal bands which correspond to different types of particle interactions ($LY\,{\approx}\,1$ for electron and $LY\,{\lesssim}\,0.2$ for nuclear recoils). The nuclear-recoil bands of the data recorded during ${\sim}1$ week of beam time (${\sim}\,5\cdot 10^5$ pulses) are shown in Fig.\,\ref{fig:2d_histogram} (2-d histogram). From kinematics using ${\sim}11$\,MeV neutrons as probes the O-recoil band extends up to ${\sim}2.4$\,MeV while the Ca- and W-bands extend up to ${\sim}1.05$\,MeV and ${\sim}240$\,keV, respectively \cite{Jagemann2005245}. Despite the strong overlap of the 3 nuclear-recoil bands the contributions of O and Ca fitted by two gaussians can be disentangled at $E_r\,{\gtrsim}\,350$\,keV (see Fig.\,\ref{fig:energyDependence} top) due to high statistics and a good light-detector resolution.
\subsection{Results and discussion}
\begin{figure}
\includegraphics[width=0.48\textwidth]{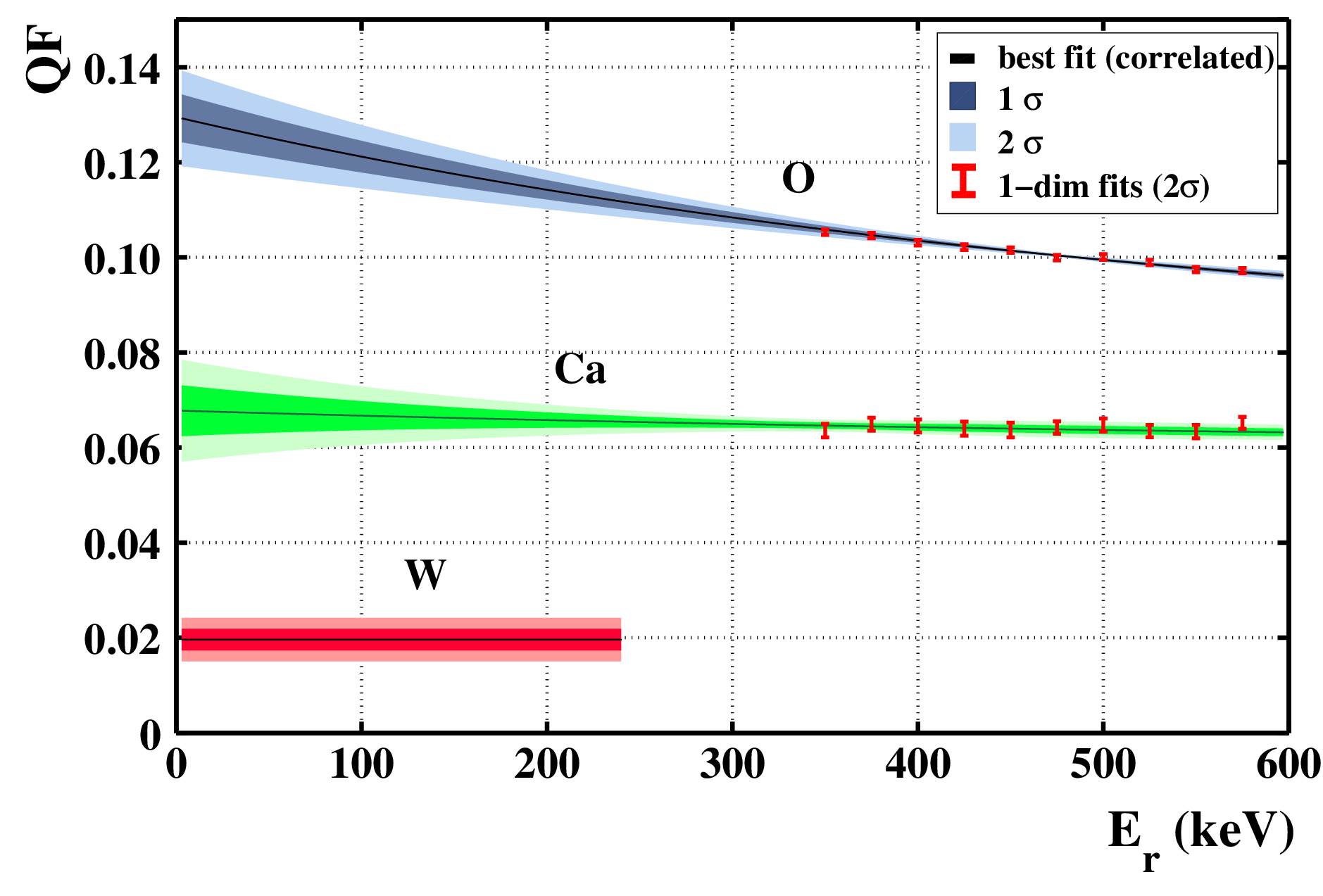} 
\caption{Results of the correlated ML analysis for $QF_O$, $QF_{Ca}$ and $QF_W$ (solid lines). The shaded areas indicate the 1$\sigma$ and 2$\sigma$ bounds.  For the first time a clear energy dependence of $QF_O$ and $QF_{Ca}$ is observed. These results are in agreement with that of the 1-dim fits of discrete energy intervals (see text) shown as red error bars. $QF_W$ is fixed (in the correlated fit) at the value measured by the triple-coincidence technique.  }
\label{fig:qf_results}
\end{figure}

 In Fig.\,\ref{fig:qf_results} the results for $QF_O$ and $QF_{Ca}$ (red error bars) derived  by these independent one-dimensional (1-dim) fits are shown for selected recoil-energy slices of 20\,keV in width. All parameters in the fit are left free except  for the LY-resolutions which are fixed  by a ML fit of the electron-recoil band \cite{strauss_PhD}. While $QF_O$ clearly rises towards lower recoil energies, this effect is less pronounced for $QF_{Ca}$.\\
\begin{figure}
\includegraphics[width=0.486\textwidth]{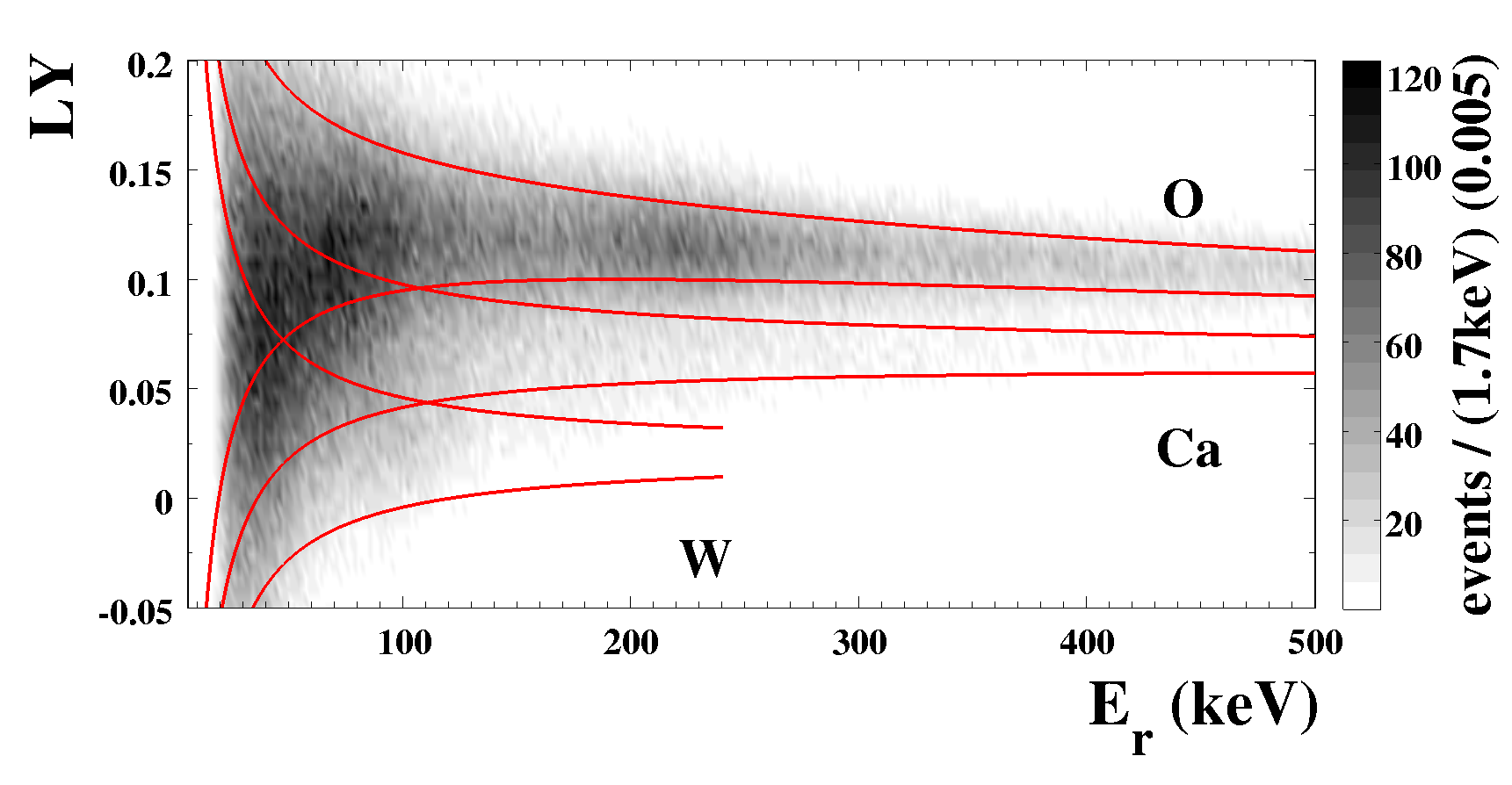}
\caption{Histogram of neutron-induced nuclear-recoil events  plotted in the LY-energy plane. The corresponding 1$\sigma$ acceptance bounds (full red lines) of O, Ca, and W as derived from the correlated ML fit (see text) are indicated.}
\label{fig:2d_histogram}
\end{figure}
\begin{figure}
\includegraphics[width=0.48\textwidth]{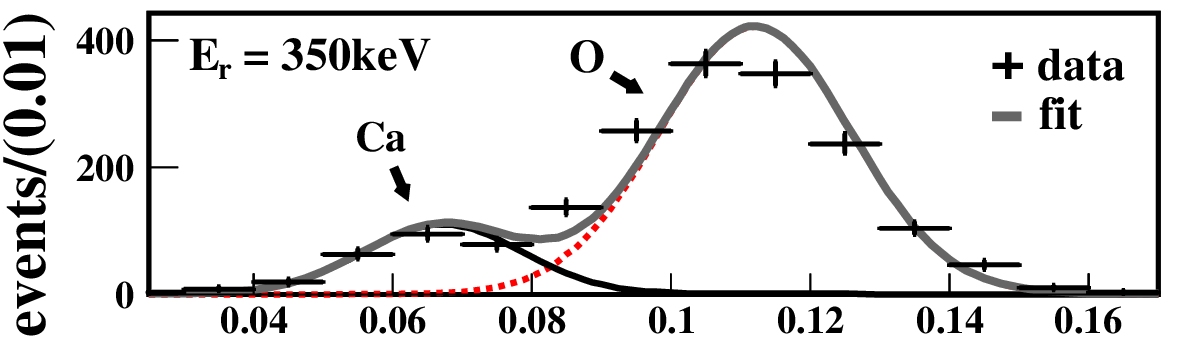} 
\includegraphics[width=0.48\textwidth]{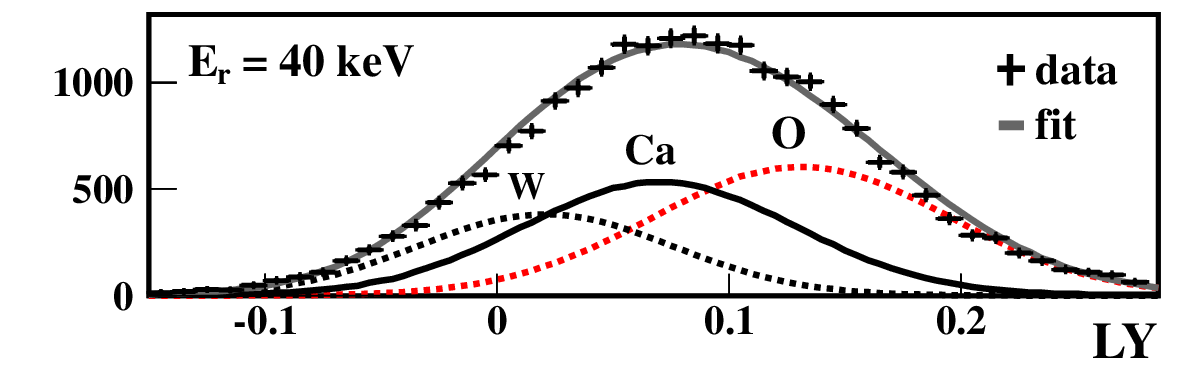}
\caption{LY histograms of energy slices (20\,keV in width) at 350\,keV (top) and 40\,keV (bottom) fitted by gaussians.   }
\label{fig:energyDependence}
\end{figure}
Below ${\sim}350$\,keV, due to the strong overlap of the nuclear recoil bands, this simple approach fails. Instead, a correlated ML fit was performed based on the\linebreak following assumptions: (1) for the mean LY of O- and Ca-scatters the phenomenological parametrization \linebreak $LY_x(E_r)\,{=}\,LY_x^\infty\left(1+f_x\cdot \exp{(-E_r/\lambda_x)}\right)$ is proposed \linebreak with the free parameters $LY_x^\infty$ (LY at $E_r\,{=}\,\infty$), $f_x$ (fraction of energy-dependent component) and $\lambda_x$ (exponential decay with energy),   and (2) the mean LY of W-scatters is approximated to be constant in the relevant energy range (up to $\sim$\,240\,keV) at the value precisely measured with the triple-coincidence technique ($LY_W\,{=}\,0.0208\,{\pm}\,0.0024$ which corresponds to\linebreak $QF_W{=}(1.96{\pm}0.22)$\%). These assumptions are supported by the result of the 1-dim fits (see Fig.\,\ref{fig:qf_results}), by Birks' model \cite{birks1964theory}, and by a recent work \cite{sabine_phD} which predict the strength of the energy-dependence to decrease with $A$. The nuclear-recoil bands are cut into energy intervals of 10\,keV (20\,keV to 1\,MeV), of 20\,keV (1\,MeV to 1.4\,MeV) and 50\,keV (above 1.4\,MeV) and fitted with up to 3 gaussians depending on the recoil energy (e.g., shown in Fig.\,\ref{fig:energyDependence} bottom for $E_r{=}40$\,keV). Except for the assumptions mentioned above and the LY-resolution all parameters are left free in the fit. The fit converges over the entire energy range (20-1800\,keV). In Table\,\ref{tab:energydependence} the results for  $LY_x^\infty$, $f_x$  and $\lambda_x$  are presented which correspond, e.g. at 40\,keV, to $QF_O{=}(12.6{\pm}0.5)$\%,  $QF_{Ca}=(6.73{\pm}0.43)$\% at $1\sigma$ C.L. Errors are dominated by systematics including different choices of the LY parametri\-zation.   The final results for $QF_O$, $QF_{Ca}$ and $QF_W$ are presented in \mbox{Fig.\,\ref{fig:qf_results}} and are found to be in perfect agreement with the outcome of the 1-dim fits (red error bars). Fig.\,\ref{fig:2d_histogram} shows the 1$\sigma$ acceptance bounds (full red lines) of O, Ca and W recoils as obtained in the correlated ML fit.\\
These are the first experimental results which clearly show a rise of $QF_O$ of ${\sim}28$\% towards the ROI (10-40\,keV) compared to that at a recoil energy of 500\,keV. For $QF_{Ca}$ the best fit yields a rise of ${\sim}6$\%, however, the energy-dependence is less significant (see Fig. \ref{fig:qf_results}). \\
\begin{table}
\caption{\label{tab:energydependence}Results for the free parameters $LY_x^\infty$, $f_x$ and $\lambda_x$ of the ML analysis. The statistical errors are given at $1\sigma$ C.L.}
\begin{tabular}{lccc}
\hline\noalign{\smallskip}
&$LY_x^\infty$&$f_x$&$\lambda_x$\\
\noalign{\smallskip}\hline\noalign{\smallskip}
O&$0.07908\pm0.00002$&$0.7088\pm0.0008$&$567.1\pm0.9$\\
Ca&$0.05949\pm0.00078$&$0.1887\pm0.0022$&$801.3\pm18.8$\\
\noalign{\smallskip}\hline
\end{tabular}
\end{table}
In previous works, the QFs of CaWO$_4$ were assumed to be constant over the entire energy range \cite{Angloher:2012vn}. A statistical analysis shows that this simple model is clearly disfavoured. Employing a likelihood-ratio test in combination with Monte-Carlo simulations gives a p-value of $p<10^{-5}$ for the data presented here to be consistent with constant QFs. Furthermore, the derived energy spectra of the individual recoiling nuclei agree with the expectation from incident 11\,MeV neutrons while the constant QF approach provides non-physical results.
\section{QFs of CRESST detectors}
In the present paper, using the 8 detector modules operated in the last CRESST measurement campaign (run32) an additional aspect was investigated: the variation of the quenching behaviour among \textit{different} CaWO$_4$ crystals \cite{strauss_PhD}. Nuclear recoils acquired during neutron calibration campaigns of CRESST run32  are completely dominated by O-scatters at  $E_r\gtrsim 150$\,keV  (from kinematics) \cite{Angloher:2012vn}. Despite low statistics (a factor of ${\sim}100$ less compared to the measurement presented here) in the available  data, the mean LY of O-events can be determined by a gaussian fit with a  precision of $\mathcal{O}$(1\%)  for every  module.   In this way, the mean QF of O between 150 and 200\,keV was determined individually for the 8 detector modules (index $i$) operated  in run32 ($\overline{QF_{O,i}^\ast}$) and for the reference detector operated at the neutron-scattering facility ($\overline{QF_O}$). Different values of $\overline{QF_{O,i}^\ast}$ are observed for the CRESST detector crystals (variation by ${\sim}11$\%) and for the reference crystal (${\sim}12$\% higher than the mean of $\overline{QF_{O,i}^\ast}$). This variation appears to be correlated with the crystal's optical quality.  The QF - which is a relative quantity -  is found to be lower if a crystal has a smaller defect density and thus a higher absolute light output, i.e., the LY of nuclear recoils is less affected by an increased defect density. This is in agreement with the prediction described in a recent work \cite{sabine_phD}. In the present paper, a simple model to account for this variation is proposed: For every detector module which is to be calibrated  a scaling factor $\epsilon_i$ is introduced, $\epsilon_i\,{=}\,\overline{QF_{O,i}^\ast}/\overline{QF_O}$. Then, within this model the QFs of the nucleus $x$ can be calculated for every module by $QF_{x,i}^\ast(E_r)\,{=}\,\epsilon_i\cdot QF_{x}(E_r)$ where $QF_{x}$ is the value precisely  measured within this work.  The nuclear-recoil behaviour of CRESST modules is well described by energy-dependent QFs. In \mbox{Table \,\ref{tab:CRESST}} the QFs, averaged over the ROI (10-40\,keV), and the scaling factor $\epsilon_i$ are listed for two selected detector modules (Rita and Daisy, with the lowest and highest absolute light output, respectively) and the mean of all 8 detector modules of run32 (\O), $QF_O^{ROI}\,{=}\,(11.2{\pm}0.5)$\%, $QF_{Ca}^{ROI}\,{=}\,(5.94{\pm}0.49)$\% and $QF_W^{ROI}\,{=}\,(1.72{\pm}0.21)$\%.\\
\begin{table}
\caption{\label{tab:CRESST} QF results averaged over the ROI (10-40\,keV) and adjusted by the scaling factor $\epsilon_i$ for the  modules Rita and Daisy, and the mean (\O) of all run32 detectors  ($1\sigma$ errors). }
\begin{tabular}{lcccc}
\hline\noalign{\smallskip}
&$\epsilon_i$&$QF_O^\mathrm{ROI}$[\%]&$QF_{Ca}^\mathrm{ROI}$[\%]&$QF_W^\mathrm{ROI}$[\%]\\
\noalign{\smallskip}\hline\noalign{\smallskip}
Rita&$0.844$&$10.8{\pm}0.5$&$5.70{\pm}0.44$&$1.65{\pm}0.19$\\
Daisy&$0.939$&$12.0{\pm}0.7$&$6.33{\pm}0.58$&$1.84{\pm}0.24$\\
\O&$0.880$&$11.2{\pm}0.5$&$5.94{\pm}0.49$&$1.72{\pm}0.21$\\
\noalign{\smallskip}\hline
\end{tabular}
\end{table}
\section{Re-analysis of latest CRESST results}
We now turn to the effect of energy-dependent quenching since constant QFs as assumed in earlier CRESST publications do not sufficiently describe the behaviour of the nuclear-recoil bands. The value of $QF_O$ in the ROI  was  underestimated by ${\sim}8$\% while the room temperature measurements overestimated the values of \linebreak $QF_{Ca}$ and $QF_W$  by ${\sim}7$\% and ${\sim}130$\%, respectively \cite{Angloher:2012vn}. Therefore, the parameter space of accepted nuclear recoils is larger than assumed in earlier publications (by ${\sim}46$\%) requiring a re-analysis of the published CRESST data.\\
During the latest measuring campaign (run32) a statistically significant signal ($4.2\sigma$) above known backgrounds was observed. If interpreted as induced by DM particles two WIMP solutions were found \cite{Angloher:2012vn}, e.g. at a mass of $m_\chi=11.6\,$GeV/c$^2$ with a WIMP-nucleon cross section of $\sigma_\chi=3.7\cdot10^{-5}$\,pb. The dedicated ML analysis was repeated using the new QF values ($\O$ in Table \ref{tab:CRESST}) yielding $m_\chi=12.0$\,GeV/c$^2$ and $\sigma_\chi=3.2\cdot10^{-5}$\,pb at 3.9$\sigma$.  Beside this moderate change of the WIMP parameters also the background composition ($e^-$, $\gamma$, neutrons, $\alpha$'s and $^{206}$Pb) is  influenced. This is mainly due to the significantly lower value of $QF_W$ which increases the leakage of $^{206}$Pb recoils into the ROI (by ${\sim}$18\%). The other WIMP solution is influenced similarly: $m_\chi$ changes  from 25.3 to 25.5\,GeV/c$^2$, $\sigma_\chi$ from $1.6\cdot10^{-6}$ to $1.5\cdot10^{-6}$\,pb and the significance drops slightly from 4.7 to 4.3$\sigma$.
\section{Summary and Outlook}
In conclusion,  the first precise measurement of $QF_W$ at mK temperatures and under conditions comparable to that of the CRESST experiment was obtained at the neutron-scattering facility in Garching by an extensive triple-coincidence technique. Furthermore, the QFs of O and Ca were precisely determined by a dedicated maximum-likelihood analysis over the entire energy range (${\sim}20{-}1800$\,keV). The observed energy dependence of the QFs, which is more pronounced for lighter nuclei, has significant influence on the determination of the ROI for DM search. Analysing CRESST neutron-calibration data a variation of the QFs between different CaWO$_4$  crystals was observed which is related to the optical quality. By the simple model proposed above the measured QFs can be adapted to every individual crystal. The updated values of the QFs are highly relevant to disentangle the recoil composition (O, Ca and W) of a possible DM signal and, therefore, to determine the WIMP parameters. Since the separation between the O and W recoil bands is higher by ${\sim}46$\% compared to earlier assumptions, background neutrons which are mainly visible as O-scatters \cite{scholl_paper} can be discriminated more efficiently from possible WIMP-induced events. A reanalysis of the run32 data shows a moderate influence of the new QF values on the WIMP parameters.\\
The results obtained here are of importance for the current CRESST run (run33) and upcoming measuring campaigns. Providing a highly improved background level run33 has the potential to clarify the origin of the observed excess signal and to set competitive limits for the spin-independent WIMP-nucleon cross section in the near future.\\
For the planned multi-material DM experiment EURECA (European Underground Rare Event Calorimeter Array) \cite{Angloher201441} the neutron-scattering facility will be an important tool to investigate the light quenching of alternative target materials in the future. 

\begin{acknowledgements}
This research was supported by the DFG cluster of excellence: “Origin and Structure of the Universe”, the DFG “Transregio 27: Neutrinos and Beyond”, the “Helmholtz Alliance for Astroparticle Phyiscs”, the “Maier Leibnitz Laboratorium” (Garching) and by the BMBF: Project 05A11WOC EURECA-XENON.
\end{acknowledgements}

\bibliographystyle{spphys}       
\bibliography{quenching_bibtex}   

%
%

\end{document}